\documentclass[11pt,a4paper]{article}
\usepackage{jheppub}

\newcommand{\Slash}[1]{{\ooalign{\hfil/\hfil\crcr$#1$}}} 

\def\lsim{\raise0.3ex\hbox{$\;<$\kern-0.75em\raise-1.1ex
\hbox{$\sim\;$}}}
\def\gsim{\raise0.3ex\hbox{$\;>$\kern-0.75em\raise-1.1ex
\hbox{$\sim\;$}}}

\title{Testing Nonstandard Neutrino Properties with a M\"ossbauer 
Oscillation Experiment}

\author[a,b]{P.A.N. Machado,}
\author[c]{H. Nunokawa,} 
\author[c]{F.~A.~Pereira dos Santos}
\author[a]{and R. Zukanovich Funchal}

\affiliation[a]{Instituto de F\'{\i}sica, Universidade de S\~ao Paulo\\
 S\~ao Paulo, C.P. 66.318, 05315-970 S\~ao Paulo, Brazil}
\affiliation[b]{Institut de Physique Th\'{e}orique, CEA-Saclay, 91191 Gif-sur-Yvette, France}
\affiliation[c]{Departamento de F\'{\i}sica, Pontif\'{\i}cia Universidade Cat\'olica do Rio de Janeiro \\
C.P. 38071, 22452-970, Rio de Janeiro, Brazil}

\emailAdd{accioly@fma.if.usp.br}
\emailAdd{nunokawa@puc-rio.br}
\emailAdd{fabio.alex@fis.puc-rio.br}
\emailAdd{zukanov@if.usp.br}

\abstract{If the neutrino analogue of the M\"ossbauer effect, namely, recoiless emission and resonant
capture of neutrinos is realized, one can study neutrino
oscillations with much shorter baselines and smaller source/detector size
when compared to conventional experiments.  
In this work, we discuss the potential of such a M\"ossbauer 
neutrino oscillation experiment to probe nonstandard
neutrino properties coming from some new physics beyond 
the standard model.
We investigate four scenarios for such new physics that modify the standard
oscillation pattern.  
We consider the existence of a light sterile neutrino that
can mix with $\bar \nu_e$, the existence of a Kaluza-Klein tower of
sterile neutrinos that can mix with the flavor neutrinos in a model
with large flat extra dimensions, neutrino oscillations with
nonstandard quantum decoherence and mass varying neutrinos,
and discuss to which extent one can constrain these scenarios.
We also discuss the impact of such new physics on 
the determination of the standard oscillation parameters.}


\begin{document}
\maketitle
\flushbottom

\section{Introduction}
\label{sec:intro}
Shortly after the discovery of the M\"ossbauer effect~\cite{Mossbauer},  
the resonant and recoil-free emission and absorption of photons by 
atoms bound in a crystal, Visscher~\cite{Visscher} suggested that 
neutrinos could also be emitted and absorbed in a similar fashion.
In the early eighties, Kells and Schiffer~\cite{Kells-Schiffer} 
proposed that a bound-state beta decay~\cite{bound-state-beta-decay, bound-state-beta-decay-2} 
could produce a recoil free emission of antineutrinos with 
ultramonochromatic energy, necessary to accomplish the neutrino 
M\"ossbauer effect. 
Such monochromatic antineutrino could be resonantly absorbed 
by an induced orbital electron capture~\cite{Mikaelyan-etal}.

More recently, in 2005, Raghavan~\cite{Raghavan, Raghavan-a} rekindle 
this idea studying the possibility for the recoilless 
$\bar{\nu}_e$ emission by the bound-state beta 
decay~\cite{bound-state-beta-decay, bound-state-beta-decay-2}
\begin{equation}
^3\text{H} \to\ ^3\text{He} + e^- \text{(bound)} + \bar{\nu}_e, 
\end{equation}
producing a $\bar \nu_e$ with energy $E = 18.6$ keV,
and the subsequent resonant $\bar{\nu}_e$ capture by the inverse 
reaction~\cite{Mikaelyan-etal}, 
\begin{equation}
^3\text{He} + e^- \text{(bound)}+ \bar{\nu}_e \to\ ^3\text{H},
\end{equation}
where the number of $\bar{\nu}_e$ captured can be inferred 
either by observing the subsequent decay of $^3$H or by directly 
counting the number of $^3$H atoms produced using some 
chemical technique. 
After this first study, a considerable amount of related
works~\cite{Raghavan2,Raghavan3,Raghavan4,Raghavan5,Potzel:2006ad,Minakata:2006ne,Minakata:2007tn,
  Bilenky-et-al,Bilenky:2007ii,Bilenky:2007vs,Akhmedov:2008jn,Bilenky:2008ez,Akhmedov:2008zz,
  Cohen:2008qb,Bilenky:2008dk,Potzel:2008xk,Kopp:2009fa,
  Potzel:2009qe,Bilenky:2009zz,Schiffer:2009zz,Potzel:2009pr,Bilenky:2011pk}
appeared in the last several years.

Due to the resonance nature of the detection process, it was estimated
in ref.~\cite{Raghavan, Raghavan-a} that the $\bar{\nu}_e$ absorption cross section
would be 12 orders of magnitude larger than the standard non-resonant
weak interaction cross section for the same energy. This would allow
for rather compact detectors, of mass of about a kg, or so, instead of a
ton or larger.

It was demonstrated in \cite{Minakata:2006ne} that a M\"ossbauer
neutrino experiment based on the $^3$H-$^3$He system, due to the very
low energy of $\bar \nu_e$ emitted (18.6 keV), can be used to study
neutrino oscillations driven by the mass squared difference relevant
to atmospheric neutrinos, $\Delta m^2_{31}$, with a baseline of only
$\sim$ 10 m. This experiment could provide precise measurements of
$\theta_{13}$ and $|\Delta m^2_{31}|$.

Moreover, if $\theta_{13}$ is not so small, which the recent 
T2K result~\cite{Abe:2011sj} seems to indicate, by extending the 
baseline to a few hundred meters, where the oscillation effect 
due to the solar mass squared difference $\Delta m^2_{21}$ becomes relevant, 
this experiment has the potential to determine also the neutrino mass 
hierarchy~\cite{Minakata:2007tn}, as first considered for reactor 
neutrinos~\cite{Petcov-etal,Choubey:2003qx}. 

Currently, almost all the existing neutrino data are very well
described by the standard 
three flavor massive and mixed neutrinos.
However, there are some experimental data which
favor more than three neutrino species.  Sterile neutrinos,
phenomenologically motivated by the results of LSND~\cite{LSND,Aguilar:2001ty} and
supported by MiniBooNe data~\cite{AguilarArevalo:2010wv}, seem to have
gained a new \'elan.  
The reactor antineutrino anomaly~\cite{Mention:2011rk}, discovered
recently after a new calculation of the reactor 
antineutrino fluxes~\cite{Mueller:2011nm,Huber:2011wv},   
as well as the cosmological data~\cite{Hamann:2010bk}, 
also seem to indicate the presence of light sterile neutrino(s).

In this paper, assuming that $\theta_{13}$ is not so small $(\sin^2
2\theta_{13} \gsim 0.01)$, we consider the possibility to probe
nonstandard neutrino properties coming from some new physics beyond
the standard model by M\"ossbauer neutrinos.  We will consider four
scenarios: the possible presence of a light sterile neutrino, mixing
with a Kaluza-Klein tower of sterile neutrinos in a model with large
extra dimensions (LED)~\cite{ADD,ADD1,ADD2}, nonstandard quantum decoherence
(NQD)~\cite{Lisi:2000zt}, and a model with the so called mass varying
neutrinos (MaVaN)~\cite{Fardon:2003eh,Kaplan:2004dq}.

Since the standard three neutrino flavor framework provides an excellent fit of 
almost all the experimental data we assume that new physics produce, at the most, subdominant
effects on top of the standard oscillation pattern.  Under this
assumption, one can try to detect small deviations from the standard
oscillation and study how to constrain new physics using M\"ossbauer
neutrinos, in a similar way as done in ref.~\cite{Ribeiro:2007jq} for
a future accelerator neutrino oscillation experiment using
conventional neutrino beam from pion decays.

\section{M\"ossbauer $\bar{\nu}_e$: Current Status}
\label{mossbauer}

In last several years, there have been various works 
on  M\"ossbauer $\bar{\nu}_e$, 
both from a theoretical and an experimental point 
of view~\cite{Raghavan2,Raghavan3,Raghavan4,Raghavan5,Potzel:2006ad,
Minakata:2006ne,Minakata:2007tn,Bilenky-et-al,Bilenky:2007ii,
Bilenky:2007vs,Akhmedov:2008jn,Bilenky:2008ez,Akhmedov:2008zz,
Cohen:2008qb,Bilenky:2008dk,Potzel:2008xk,Kopp:2009fa,
Bilenky:2009zz,Potzel:2009qe,Schiffer:2009zz,Potzel:2009pr,Bilenky:2011pk}. 
Let us make a brief summary of the current status
of the prospect of a M\"ossbauer $\bar{\nu}_e$ oscillation
experiment.

\subsection{Theoretical considerations on M\"ossbauer $\bar{\nu}_e$ oscillation}
\label{mossbauer-theory}

Despite that neutrino oscillations are believed to have been observed 
and confirmed experimentally, a complete consensus on 
the formalism of neutrino oscillations seems to be 
still lacking (see e.g. \cite{Akhmedov:2009rb}). 
Indeed, due to the very special nature of M\"ossbauer $\bar{\nu}_e$,
there has been some controversy in the literature whether or not 
M\"ossbauer $\bar{\nu}_e$ would indeed oscillate~\cite{Bilenky:2008ez,Bilenky:2008dk,Akhmedov:2008jn,Akhmedov:2008zz, Cohen:2008qb,Bilenky:2009zz}.

It was argued in refs.~\cite{Bilenky:2008ez,Bilenky:2008dk,Bilenky:2011pk} 
that the very small energy uncertainty on M\"ossbauer $\bar{\nu}_e$ 
(due to its ultramonochromatic nature) is in conflict with the 
energy uncertainty required to observe neutrino oscillations. In this case, 
M\"ossbauer neutrinos could be used to test different approaches on the 
formalism of neutrino oscillations.

In ref.~\cite{Akhmedov:2008jn}
the oscillation probability of M\"ossbauer $\bar{\nu}_e$
was calculated based solely on quantum field theory 
without making any a priori assumption about 
the energy and momentum of the intermediate
neutrino state. 
It was concluded that despite the nearly perfect monochromaticity of the beam, 
M\"ossbauer neutrinos do oscillate (see also \cite{Kopp:2009fa}). 
The same conclusion was also drawn in ref.~\cite{Cohen:2008qb}. 

In this work we assume that M\"ossbauer $\bar{\nu}_e$
do oscillate and that the standard expression for 
three active neutrino flavors oscillation probability
can be used. We modify this expression 
accordingly to the new physics models we consider. 

\subsection{Experimental feasibility of a M\"ossbauer $\bar{\nu}_e$ experiment}
\label{mossbauer-exp}

The natural line width of a $\bar{\nu}_e$ from a $^3$H 
(with life time $\tau = 17.8$ yr) decay is 
$\Gamma = \hbar/\tau \simeq 1.17 \times 10^{-24}$ eV and 
if there is no recoil, this implies the extremely 
small energy uncertainty, $\Delta E/E \sim 10^{-31}$,
which, however, is impossible to reach experimentally. 
In order to prevent recoil, Raghavan~\cite{Raghavan, Raghavan-a} 
considered that both $^3$H and $^3$He should be embedded in Nb 
metal lattices, and estimated that, due to several line broadening effects, 
the relative energy uncertainty would be $\Delta E/E \sim 5\times 10^{-16}$, 
implying a resonant capture cross section of the order of 
$\sim 10^{-33}$ cm$^2$. 

If such a large value of the cross section can be realized, 
in the absence of oscillation, about one million events per day  
would be expected for 1 MCi source and 100 g $^3$He target at 
a baseline of $\sim$ 10 m. 
However, in ~\cite{Potzel:2006ad}, it was argued that 
this value could be significantly reduced by some other 
line broadening effects missed in the estimation done in ~\cite{Raghavan,Raghavan-a}. 

More recently, it was claimed in ref.~\cite{Raghavan2,Raghavan3,Raghavan4,Raghavan5} that, 
due to motional averaging by lattice vibrations, 
the decay of $^3$H in crystals can emit a hypersharp 
neutrino with $\Delta E/E \sim 5\times 10^{-29}$, 
implying a capture cross section of  $\sim 10^{-17}$ cm$^2$. 
This conclusion was criticized by 
Refs.~\cite{Schiffer:2009zz,Potzel:2009qe,Potzel:2009pr}
claiming it would be impossible to reach such a value 
and that it would be impractical to perform the experiment using 
the $^3$H-$^3$He system. 
For example, it was stressed~\cite{Schiffer:2009zz,Potzel:2009pr}
that $^3$H and $^3$He atoms occupy differently the lattice space, 
implying some energy difference (by lattice expansion or 
contraction) before and after the emission and absorption of $\bar{\nu}_e$, 
which would broaden the natural line width by many orders of magnitude. 

In ref.~\cite{Potzel:2009pr} it was proposed that another system,
$^{163}$Ho-$^{163}$Dy, would be more promising than the $^3$H-$^3$He one.
We note that this new system would imply an even smaller baseline $\lsim 2$ m 
in order to study $\Delta m^2_{31}$ driven-oscillations, due to a lower 
$\bar{\nu}_e$ energy, $E = 2.6$ keV.

While it is yet far from clear if a M\"ossbauer $\bar{\nu}_e$
experiment can be really realized, we assume that it will become
possible in the future and for definiteness, throughout this work, we
consider the $^3$H-$^3$He system as $\bar{\nu}_e$ emitter/absorber
with $E = 18.6$ keV.  We note, however, that our analysis method can
be applied to other systems by appropriately re-scaling neutrino
energies and baselines.

\section{On the Framework and Assumptions on New Physics}
\label{framework}

In this section we describe the framework as well as the 
assumptions for the new physics to be probed 
by a M\"ossbauer neutrino experiment. 
Since all of these new physics models are already described in detail
in previous works, we will provide only a brief descriptions of each
model and refer the readers to the appropriate references in each 
case.

As mentioned in the introduction, most of the experimental data
are well described by the standard three flavor oscillation scheme, 
allowing us to assume that the effect coming from new physics is small
(subdominant).
Therefore, throughout this work, even in the presence of new
physics, we consider, to a good approximation, the following 
true (input) values of the standard oscillation parameters 
determined by the three flavor analysis of experimental data:
$\Delta m_{21}^2 = 7.6 \times 10^{-5}$ eV$^2$, $\sin^2 \theta_{12}$ = 0.31, 
$|\Delta m_{31}^2| = 2.4 \times 10^{-3}$ eV$^2$
where the mass squared differences are defined as $\Delta m^2_{ij}
\equiv m^2_i - m^2_j$ with $m_i\ (i=1,2,3)$ being the neutrino mass.
For the most recent global analyses of the neutrino oscillation data which have 
taken into account the new T2K result~\cite{Abe:2011sj} as well as new 
calculations of the reactor neutrino fluxes~\cite{Mueller:2011nm,Huber:2011wv},
see Refs.~\cite{Fogli:2011qn,Schwetz:2011zk}.

As long as the mixing among the standard active three neutrino flavors
is concerned, we consider the parameterizations found in
ref.~\cite{Nakamura:2010zzi}.
We note that the values of the CP phase $\delta$ and 
of the angle $\theta_{23}$ are irrelevant for the
$\bar{\nu}_e \to \bar{\nu}_e$ channel, even in the
presence of new physics. 
We define the lightest neutrino mass $m_0$ as $m_0 = m_1\ (m_3)$ for
normal (inverted) mass hierarchy.  As we will see, unlike the 
standard oscillation case, for LED and MaVaN, the oscillation probabilities
depend also on the absolute neutrino mass scale $m_0$.

\subsection{A Light Sterile Neutrino}
\label{steril-framework}

The original motivation for considering a light sterile neutrino was 
the result of the LSND experiment~\cite{LSND,Aguilar:2001ty}, now also supported 
by MiniBooNe~\cite{AguilarArevalo:2010wv}, where the data can be 
interpreted  as oscillation between active and 
sterile neutrinos with a mass squared difference of 
$\sim 0.1-1$ eV$^2$. 
Another hint in favor of a light sterile neutrino comes from the
GALLEX~\cite{GALLEX-Source,Kaether:2010ag} and SAGE~\cite{SAGE-Source} $^{51}$Cr
neutrino source experiments. Both measured a deficit of $\nu_e$
events with respect to the prediction. This can be a signal of 
oscillation from active to sterile neutrinos~\cite{Giunti-etal-Ga,Giunti:2010wz}.

More recently, the so called reactor antineutrino anomaly~\cite{Mention:2011rk}
supports also the possibility of oscillation to a sterile neutrino driven by 
a mass squared difference compatible with LSND and MiniBooNe. 
In addition, though the significance is not yet strong, cosmological
data also favors the presence of sub-eV mass sterile
neutrinos~\cite{Hamann:2010bk}.

We note, however, that the significance of the LSND excess was diminished from 
3.8 to 2.9 $\sigma$ according to the new result on pion production from the HARP-CDP 
collaboration~\cite{Bolshakova:2011hr}, and  more recent MiniBOONE result, based 
on the $8.58 \times 10^{20}$ POT, also reduced the significance of 
the $\bar{\nu}_\mu \to \bar{\nu}_e$ excess to 0.84 $\sigma$~\cite{MiniBOONE-Nufact11}.

Oscillation between active and sterile neutrinos can be tested using
the $\bar{\nu}_e$ disappearance mode~\cite{Kopeikin:2003uu} in reactor
neutrino experiments.  In ref.~\cite{deGouvea:2008qk} the impact of
sterile neutrinos on the determination of the standard oscillation
parameters $\theta_{13}$ and $\Delta m^2_{31}$ for reactor neutrinos 
was studied.

Here we consider the so called 3+1 model where one species of
light sterile neutrino is added to the standard 3 flavor framework.
See ref.~\cite{sterile3+1,Maltoni:2002xd,Sorel:2003hf,Maltoni:2007zf,Karagiorgi:2009nb,Kopp:2011qd,Giunti-laveder:2011} for a partial list of works that studied
this possibility. In this model, the mixing between 4 neutrinos (3 active
and 1 sterile) is described by six mixing angles and 3 CP
phases.
For simplicity, we take only one of the mixings, the one which  
involves the 4th mass eigenstate (mainly the sterile neutrino state), 
$\theta_{14}$,  different from zero.   

For definiteness, we consider the mixing between active and sterile 
neutrinos as 
\begin{eqnarray} 
\left( \begin{array}{c} 
                   \nu_e \\ \nu_\mu \\ \nu_\tau \\ \nu_s
                   \end{array}  \right)
 = U
\left( \begin{array}{c} 
                   \nu_1 \\ \nu_2 \\ \nu_3  \\ \nu_4 
                   \end{array}  \right)\, ,
\label{mixing-steril}
\end{eqnarray}
where 
\begin{eqnarray}
U = 
 \left[ 
                   \begin{array}{cccc}
               c_{14}  &  0 & 0 & s_{14}  \\
                   0  &  1 & 0 & 0       \\
                   0  &  0 & 1 & 0   \\
             -s_{14}  &  0 & 0 & c_{14}
                   \end{array} \right]
 \left[ 
                   \begin{array}{cccc}
                1 &  0   & 0 & 0   \\
                 0  &  c_{23} & s_{23} & 0       \\
                   0  &  -s_{23} & c_{23} & 0   \\
             0  &  0 & 0 & 1
                   \end{array} \right] 
  \left[ 
                   \begin{array}{cccc}
                 c_{13} &  0   &  s_{13} & 0   \\
                 0  & 1 & 0 & 0       \\
                   -s_{13}  &  0 & c_{13} & 0   \\
             0  &  0 & 0 & 1
                   \end{array} \right]
 \left[ 
                   \begin{array}{cccc}
                 c_{12} &  s_{12}   &  0 & 0   \\
                  -s_{12}  & c_{12} & 0 & 0       \\
                  0  &  0 & 1 & 0   \\
             0  &  0 & 0 & 1
                   \end{array} \right],
\end{eqnarray}
with the notation $c_{ij} \equiv \cos \theta_{ij}$ and $s_{ij} \equiv \sin \theta_{ij}$. Since CP violation is not observable in the $\bar{\nu}_e \to
\bar{\nu}_e$ channel, we ignore all CP phases.

\begin{figure}[!h]
\begin{center}
\hglue -0.2cm
\includegraphics[width=0.8\textwidth]{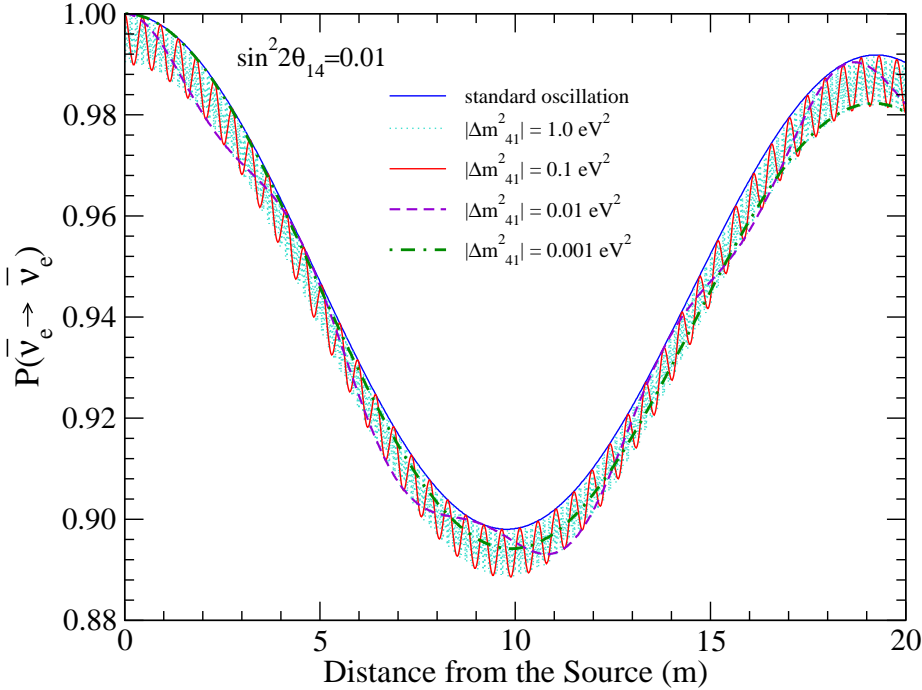}
\end{center}
\vglue -0.5cm
\caption{
$\bar{\nu}_e$ survival probability 
as a function of the distance from the source 
for the 3+1 model, the standard 3 active flavors 
plus one light sterile neutrino, and $E=18.6$ keV. 
We set the mixing angle between active and sterile as 
$\sin^2 2 \theta_{14} = 0.01$ ($\theta_{i4} = 0$ for $i\ne 1$) 
and $| \Delta m^2_{41} |  = 1, 0.1, 0.01$ and 0.001 eV$^2$. 
Here $\sin^2 2\theta_{13}$ = 0.1.
For the purpose of comparison, the probability for the standard 
oscillation scenario without a sterile neutrino is also shown 
by the solid blue curve. }
\label{fig-probab-steril}
\end{figure}

Under this parameterization, vacuum oscillation probabilities 
can be easily calculated without any approximation.
In figure~\ref{fig-probab-steril} we show the $\bar{\nu}_e$ survival
probability as a function of the baseline for the 3+1 model, for 
$\sin^2 2\theta_{14} = 0.01$ and $| \Delta m^2_{41}| =
1, 0.1, 0.01$ and 0.001 eV$^2$.  Assuming that the mixing angle
$\theta_{14}$ is small, we see that in this range of distances the presence 
of a sterile neutrino induces an extra smaller {\it modulation} on top of 
the standard oscillation pattern. We
note that for larger value of $\Delta m^2_{41}$, the net effect is
expected to be similar to that of LED to be discussed in
section \ref{LED-framework}.

\subsection{Large Extra Dimensions}
\label{LED-framework}

We consider the model of large extra dimensions discussed in
\cite{Barbieri:2000mg,Davoudiasl:2002fq,Mohapatra-et-al,Mohapatra2,Mohapatra3,Machado:2011jt,Machado:2011kt}
in connection with neutrino physics, based on the so called flat large
extra dimension (LED) scenario~\cite{ADD,ADD1,ADD2}.  In this model, it is
assumed that right handed neutrinos (Standard Model singlet fields)
can, as well as gravity, propagate in the $d$-dimensional bulk, while
Standard Model (SM) particles can only propagate in a brane of 3+1
dimensions.  While LED induced neutrino oscillations is not favored by
most of the neutrino data~\cite{Machado:2011jt}, in
ref.~\cite{Machado:2011kt} it was demonstrated that
gallium~\cite{GALLEX-Source,Kaether:2010ag,SAGE-Source,Giunti-etal-Ga,Giunti:2010wz} and reactor
antineutrino~\cite{Mention:2011rk} anomalies can be explained by this
scenario.

As in~\cite{Machado:2011jt}, we do not consider explicitly  
how many extra spatial dimensions  do exist, but we assume that 
the largest one, compactified on a torus of radius $a$, is sufficiently 
larger than the others so effectively only 4+1 dimensions can be considered.
In other words, only the largest LED in practice contribute to modify the
oscillation probabilities.  Since in any case for us the LED effect is
a subdominant one in neutrino oscillations, this assumption looks reasonable.

For this effective model, the 4-dimensional Lagrangian which describes
the charged current interaction of the brane neutrinos with the {\it
  W} as well as the mass term resulting from these couplings with the
bulk fermions in the brane, after electroweak symmetry breaking and
dimensional reduction, can be written as~\cite{Davoudiasl:2002fq},

\begin{eqnarray}
\hskip -0.5cm
\mathcal{L^{\text{eff}}_{\text{LED}}} \, &= &\, \mathcal{L_{\text{mass}}} +
\mathcal{L_{\text{CC}}} \nonumber \\ &= & \displaystyle
\sum_{\alpha,\beta}m_{\alpha\beta}^{D}\left[\overline{\nu}_{\alpha
    L}^{\left(0\right)}\,\nu_{\beta R}^{\left(0\right)}+\sqrt{2}\,
  \sum_{N=1}^{\infty}\overline{\nu}_{\alpha
    L}^{\left(0\right)}\,\nu_{\beta
    R}^{\left(N\right)}\right]  
 +  \sum_{\alpha}\sum_{N=1}^{\infty}\displaystyle
\frac{N}{a}\, \overline{\nu}_{\alpha L}^{\left(N\right)} \,
\nu_{\alpha R}^{\left(N\right)}\nonumber \\ 
&+&  \displaystyle
\frac{g}{\sqrt{2}}\,\sum_{\alpha} \,
\overline{l_\alpha}\gamma^{\mu}\left(1-\gamma_{5}\right)\nu_{\alpha}^{\left(0\right)}\,
W_{\mu}+\mbox{h.c.},
\end{eqnarray}
where the Greek indices $\alpha,\beta = e,\mu,\tau$, the capital Roman
index $N=1,2,3,...,\infty$, $m_{\alpha \beta}^{D}$ is a Dirac mass
matrix, $\nu^{(0)}_{\alpha R}$, $\nu^{(N)}_{\alpha R}$ and
$\nu^{(N)}_{\alpha L}$ are the linear combinations of the bulk fermion
fields that couple to the SM neutrinos $\nu^{(0)}_{\alpha L}$
which is identified, from now on,  
as $\nu_{\alpha}\ (\alpha = e,\mu, \tau$) for simplicity. 

After performing unitary transformations in order to diagonalize
$m^{D}_{\alpha \beta}$ we arrive at the neutrino evolution equation
(see eq. (A7) of ref.~\cite{Machado:2011jt}) 
that can be solved to obtain the eigenvalues $\lambda_j^{(N)}$ 
and amplitudes $W_{ij}^{(0N)}$ (see Appendix of ref.~\cite{Machado:2011jt}). 

Then the $\bar{\nu}_e$ survival probability 
at a distance $L$ from production 

\begin{equation}
P(\bar{\nu}_e \to \bar{\nu}_e;L) = 
\vert {\cal{A}}(\bar{\nu}_e \to \bar{\nu}_e; L)\vert^2 \, ,
\end{equation}
can be given in terms of the transition amplitude
\begin{eqnarray}
{\cal{A}}(\bar{\nu}_e \to \bar{\nu}_e; L)  &= & 
\displaystyle \sum_{i,j,k=1}^{3}\sum_{N=0}^{\infty} U_{e i} U_{e k}^{*}
W_{ij}^{(0N)*}W_{kj}^{(0N)} 
\times \exp \left(i\frac{\lambda_j^{(N)2}L}{2Ea^2} \right)\, ,
\label{eq:amplitude}
\end{eqnarray}
where $E$ is the neutrino energy, $L$ is the baseline distance, 
$a$ is the size of the largest extra dimension, 
$U$ and $W$ are the mixing matrices for active and KK 
(Kaluza-Klein) neutrino modes,
respectively (see \cite{Machado:2011jt}), 
$\lambda_j^{(N)}$ are the eigenvalues of the Hamiltonian matrix 
described in eq. (A11) in \cite{Machado:2011jt},
and the index $N$ refers to the KK modes.

\begin{figure}[!h]
\begin{center}
\hglue -0.2cm
\includegraphics[width=0.8\textwidth]{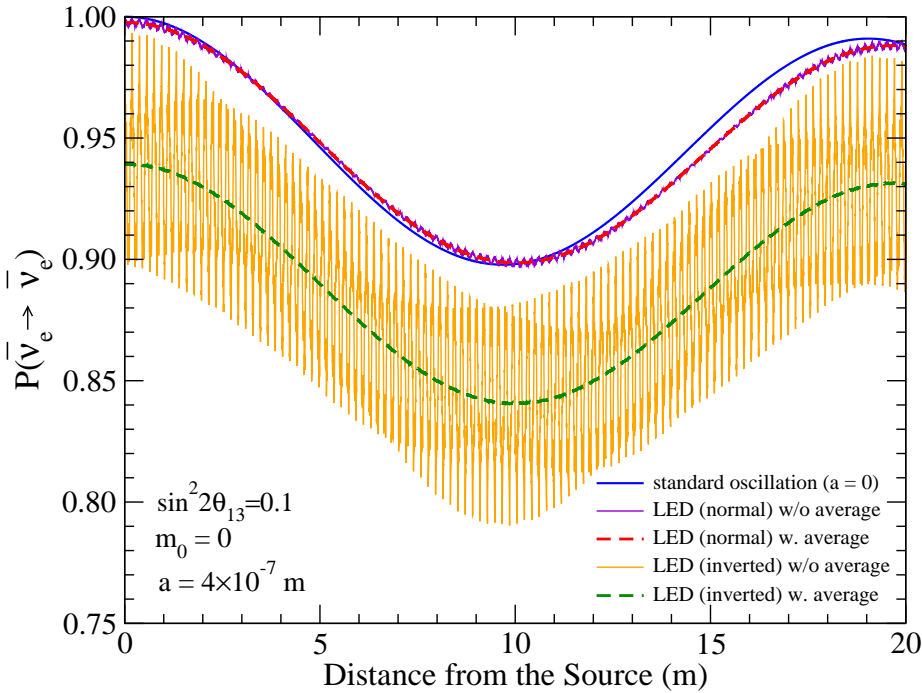}
\end{center}
\vglue -0.5cm
\caption{
$\bar{\nu}_e$ survival probability
as a function of the distance from the source 
for the LED model with $a =0.4$ $\mu$m and the lightest neutrino mass 
set to zero for the normal ($m_1 = m_0=0$) 
 (solid magenta curve) and 
inverted ($m_3 = m_0 = 0$) hierarchy (solid orange curve). 
Here $\sin^2 2\theta_{13}$ = 0.1.
We also show the case where the probabilities are averaged 
over the production and detection (interaction) 
points using the Gaussian smearing function described in 
eq.~(\ref{eq:gaussian}) of the Appendix with 
$\sigma_L $ = 10 cm, 
by the dashed red (normal) and green (inverted) curves. 
For the purpose of comparison, the standard 
survival probability without LED  is shown by the solid 
blue curve. 
}
\label{fig-probab-led}
\end{figure}

In figure~\ref{fig-probab-led} we show an example of 
the $\bar{\nu}_e$ survival probability with the effect of LED, 
for $a=0.4$ $\mu$m and  both mass hierarchies.
As discussed
in~\cite{Barbieri:2000mg,Davoudiasl:2002fq,Mohapatra-et-al,Mohapatra2,Mohapatra3,Machado:2011jt},
the presence of LED induces conversion from active to sterile KK mode
neutrinos with rapid oscillations (or smaller oscillation lengths) and
reduce further the $\bar{\nu}_e$ survival probability when compared to
the standard oscillation without LED.  In addition to the
overall reduction of the probability, LED induces some shift
(distortion) of the oscillation minimum though this effect is not so
large.

In agreement with the behavior of the $\bar{\nu}_e$ survival
probability for reactor neutrinos shown in figure 1 of
ref.~\cite{Machado:2011jt}, for a given value of $a$, the impact of
LED is significantly larger for the case of the inverted hierarchy
(orange curve) than that of the normal one (purple curve).
The reason why the LED effect is larger for the inverted hierarchy 
is that for the normal one, there is a suppression due to  
small $\theta_{13}$~\cite{Machado:2011jt}. 

Due to the ultra monochromatic energy, in principle,  a
M\"ossbauer neutrino experiment could be highly sensitive 
to the LED effect. 
However, due to the uncertainties 
on the exact production and detection points, {\it i.e.}  
the finite size of the source and detector, the large LED effect 
can be significantly washed out. This is exemplified 
by the dashed curves shown in figure~\ref{fig-probab-led}.

\subsection{Nonstandard Quantum Decoherence}
\label{decoherence-framework}

Even in the absence of new physics, loss of coherence can 
occur in standard neutrino oscillations if neutrinos travel further 
than the coherence length (see e.g.\cite{Giunti:1991ca}). This, in fact,  
can be important for neutrinos from astrophysical sources 
traveling very long distances. 
Decoherence can also happen in dense media when 
 collisions become important (see e.g. \cite{Raffelt:1996wa}), 
in particular, when neutrino-neutrino interactions are significant
(see e.g. \cite{Raffelt:2007yz}),  
or matter density perturbations/fluctuations are 
relevant~\cite{Fogli:2006xy}.
Here we focus on a different kind of decoherence, what we will refer to as
nonstandard quantum decoherence (NQD), a decoherence that could be induced by 
quantum gravity~\cite{ellis}.

We assume that the survival probability of $\bar \nu_e$ 
in the presence of the nonstandard decoherence effect in the 1-3 sector, 
is given by~\cite{Lisi:2000zt,Gago:2000qc},
\begin{eqnarray}
&& \hskip -0.8cm 
P ( \overline{\nu}_e \rightarrow \overline{\nu}_e )  
 = 1- 
{1\over 2}\sin^2 2 \theta_{13}
\left[ 1 - e^{-\gamma(E)L} \cos (\Delta_{31})\right]
+ P_\odot,
\label{eq:prob-dc}
\end{eqnarray}
where $\Delta_{31} \equiv \Delta m_{31}^2L/(2E)$ and 
$P_\odot$ is the part of the probability that depends 
on the solar oscillation parameters, 
\begin{equation}
\hskip -0.2cm
P_\odot \equiv 
\frac{1}{2} s^2_{12} \sin^2 2\theta_{13} 
\sin(\Delta_{31}) 
\sin(\Delta_{21})  + 
 \left[c_{13}^4 \sin^2 2\theta_{12} + 
s_{12}^2 \sin^22 \theta_{13} \cos (\Delta_{31}) 
\right] \sin^2 \left(\frac{\Delta_{31}}{2}\right),
\end{equation}
where $\Delta_{21} \equiv \Delta m_{21}^2L/(2E)$ 
and 
$c_{ij} \equiv \cos\theta_{ij}$ and $s_{ij} \equiv \sin\theta_{ij}$.
Strictly speaking, there should be some interference term which 
depends on both, the decoherence parameters and the solar parameters, but
since we consider the case where decoherence is a subdominant effect, 
we assume that such term is negligible in our case. 

As in previous works~\cite{Lisi:2000zt,Gago:2000qc,Ribeiro:2007jq}, 
we assume that the parameter $\gamma (E)$ can be 
phenomenologically parameterized as, 
\begin{eqnarray}
\gamma (E) = \gamma_0
\left( \frac{ E }{ \text{ GeV } } \right)^\beta,
\label{E-dep}
\end{eqnarray}
where $\gamma_0$ and $\beta$ are constant.
In this work, the parameter $\beta$ is restricted to be
in the range $-2 \leq \beta \leq 2$. 
%

\begin{figure}[h]
\begin{center}
\hglue -0.4cm
\includegraphics[width=0.78\textwidth]{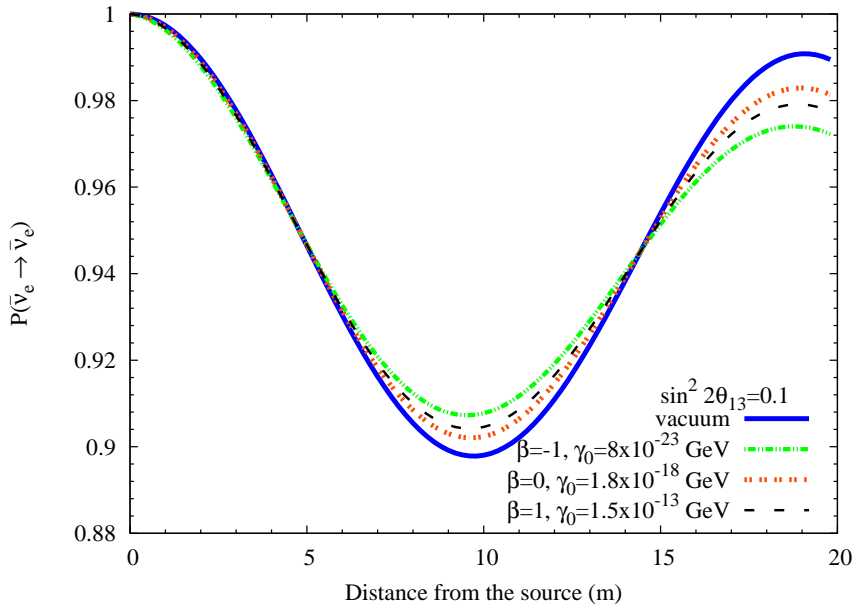}
\end{center}
\vglue -0.55cm
\caption{
$\bar{\nu}_e \to \bar{\nu}_e$ survival probability 
as a function of the distance from the source for 
the decoherence model, for the cases 
$(\gamma_0,\beta) = (8\times10^{-23}\text{ GeV},-1), 
(1.8\times 10^{-18} \text{ GeV}, 0)$ and 
$(1.5\times 10^{-13} \text{ GeV}, +1)$. 
Here $\sin^2 2\theta_{13}$ = 0.1.}
\label{fig-probab-dc}
\end{figure}

To illustrate the effect of decoherence 
in terms of probability, we show in figure~\ref{fig-probab-dc} 
how the survival probability is modified by this effect
for $\sin^2 2\theta_{13}$ = 0.1 and three
cases $(\gamma_0,\beta) = (8\times10^{-23}\text{GeV},-1), 
(1.8\times 10^{-18} \text{GeV}, 0)$ and 
$(1.5\times 10^{-13} \text{GeV}, +1)$. 
As expected, the net effect of NQD is to reduce 
the oscillation amplitudes as compared to the standard oscillation case.

\subsection{Mass Varying Neutrinos}
\label{MaVaN-framework}

Some years ago a connection between neutrino mass and dark energy was 
proposed in a scenario known as Mass Varying Neutrinos 
(MaVaN)~\cite{Fardon:2003eh}. The idea was that 
neutrino mass comes about from the interaction with a scalar field 
whose effective potential changes with the local neutrino density.
So the neutrino mass would be a dynamical variable that 
depends on the local neutrino density (therefore vary as the Universe evolves). 
Due to the connection field the dark energy density could keep track of the 
 matter densities (dark matter, baryons and neutrinos) 
throughout the evolution of the Universe. 
One could further consider that if the scalar field is in some way coupled to 
visible matter, the neutrino mass could depend on the local 
matter density as well~\cite{Kaplan:2004dq,Gu:2003er}.

One can find in the literature phenomenological studies of MaVaN
models involving
solar~\cite{Barger:2005mn,Cirelli:2005sg,GonzalezGarcia:2005xu,deHolanda:2008nn}
and atmospheric neutrinos~\cite{Abe:2008zza}.  
(See ref.~\cite{Franca:2009xp} where the cosmological impact of MaVaN was 
studied.) 
We adopt here
essentially the same framework of these references but for the 1-3
sector (mixing between the first and third generation), as discussed
in ref.~\cite{Schwetz:2005fy} for future reactor neutrino experiments.

If the effect of MaVaN is present and significant, a M\"ossbauer neutrino 
oscillation experiment should be able to 
detect some deviation of the oscillation probability, 
which will depend on the matter present between the source 
and the detector, from the standard vacuum oscillation.  
One of the advantages of a M\"ossbauer experiment is 
that it is very easy to switch on and off the matter 
induced MaVaN effect by placing (and removing) 
matter between the source and the detector 
which are separated by only $\sim O(10)$ m. 

The Lagrangian we consider has the same form 
as assumed in ref.~\cite{GonzalezGarcia:2005xu}, 
and given by 
\begin{equation}
\hskip -0.2cm
\mathcal{L^{\text{eff}}_{\text{MaVaN}}} = \displaystyle
\sum_i \bar{\nu}_i (i \Slash{\partial} - m_i)\nu_i + 
\sum_f \bar{f} (i \Slash{\partial} - m_f)f + 
\frac{1}{2}\left[\phi (\partial^2 - m^2_S) \phi \right] 
+\sum_{ij} \lambda^\nu_{ij}\bar{\nu}_i \nu_j \phi  
+ \sum_{f} \lambda^f \bar{f} f\phi,
\end{equation}
where $m_i$ $(i=1,2,3)$ are neutrino masses 
in the presence of the cosmic neutrino background, which 
are regarded as {\it vacuum} neutrino masses, $m_f$ is the 
mass of fermion of $f$-species, 
$m_S$ is the mass of the scalar particle (acceleron) 
responsible for the accelerated expansion 
of the universe (which behaves as the dark energy),  
and 
$\lambda^\nu_{ij}$ and $\lambda^{f}$ are, respectively, 
the effective neutrino-scalar and matter-scalar couplings,
and $f$ refers to fermions $e$, $n$ and $p$.

The effective neutrino evolution equation in the MaVaN scenario 
considered in this work is given by, 
\begin{eqnarray} 
\hskip -0.2cm
i {d\over dt} \left( \begin{array}{c} 
                   \nu_e \\ \nu_\mu \\ \nu_\tau 
                   \end{array}  \right)
\hskip -0.2cm
 = \frac{1}{2E} \left[ 
                  \left( \begin{array}{ccc}
                   A  &  0 & 0   \\
                   0  &  0 & 0    \\
                   0  &  0 & 0
                   \end{array} \right) 
+ U{\cal M}^2 
                   U^{\dagger} 
                   \right] 
\hskip -0.2cm
\left( \begin{array}{c} 
                   \nu_e \\ \nu_\mu \\ \nu_\tau 
                   \end{array}  \right),
\label{evolution-MavaN}
\end{eqnarray}
where $A \equiv 2\sqrt{2} G_F n_e E$, 
$G_F$ and $n_e$ are the Fermi constant and 
the electron number density, respectively, and 
the effective mass squared matrix is given by   
\begin{equation}
{\cal M}^2 \equiv 
                  \left( \begin{array}{ccc}
                   m_1^2 & 0 & M^2_{13}(r)   \\
                   0 &  m_2^2 & 0    \\
                    M^2_{31}(r)    &  0 &\{m_3-M_{33}(r)\}^2
                   \end{array} \right). 
\label{eq:mass-matrix}
\end{equation}
As in the framework considered in \cite{GonzalezGarcia:2005xu}, 
$M_{ij}$ is related to more fundamental parameters 
of the MaVaN model as 
\begin{equation}
 M_{ij}(r) = 
\frac{\lambda^\nu_{ij}}{m^2_S} 
\sum_f \lambda^f n_f(r),
\label{eq:scalar-coupling}
\end{equation}
where $n_f(r)$ is the number density of fermion
of $f$-species. 
For simplicity, we only consider 
the case of vanishing lightest neutrino
mass $m_1 = 0$ ($m_3 = 0$) for normal (inverted) 
mass hierarchy. 

Following ref.~\cite{GonzalezGarcia:2005xu}, 
we introduce the effective 
MaVaN parameters $\alpha_{13}$ and $\alpha_{33}$ 
for the 1-3 sector as
\begin{equation}
M_{ij}(r) \equiv \alpha_{ij} \left[\frac{\rho}{\text{g/cm}^3} \right],
\ \ (i,j) = (1,3), (3,3), 
\end{equation}
where $\rho$ is the density of the matter present 
along the neutrino trajectory. In vacuum 
our evolution equation coincides with the standard one. 

\begin{figure}[!h]
\begin{center}
\hglue -0.4cm
\includegraphics[width=0.8\textwidth]{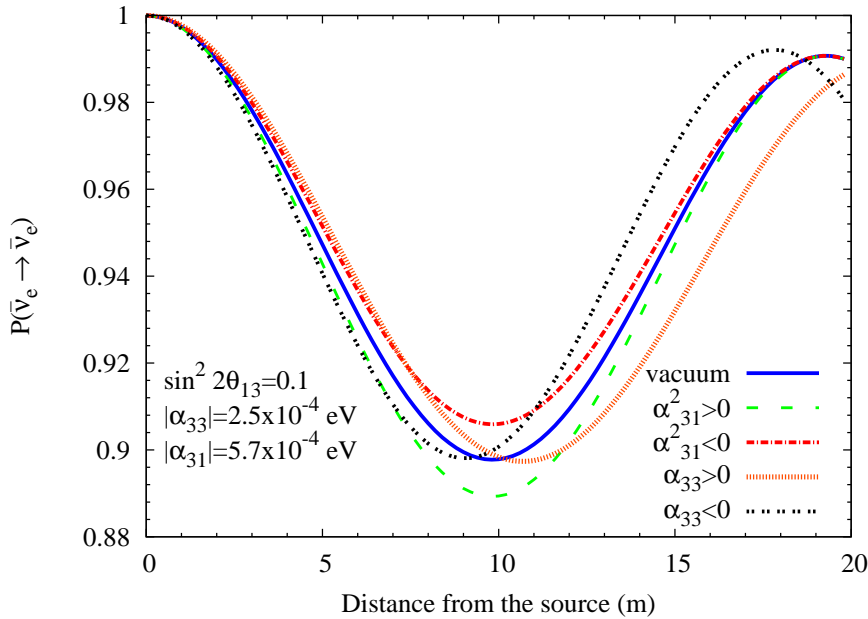}
\end{center}
\vglue -0.55cm
\caption{
$\bar{\nu}_e \to \bar{\nu}_e$ survival probability as 
a function of the distance from the source 
for the MaVaN model and four cases: 
$(\alpha_{33},\alpha_{31}) = (\pm 5.7 \times 10 ^{-4}$ eV, 0), 
$(0, 2.5 \times 10 ^{-4}$ eV) and 
$(0, 2.5i \times 10 ^{-4}$ eV). 
Here $\sin^2 2\theta_{13}$ = 0.1.
To study the MaVaN effect, we consider the case 
where we place iron (with density $\rho = 7.9$ g/cm$^3$) 
between the source and the detector.}
\label{fig-probab-mavan}
\end{figure}

In figure~\ref{fig-probab-mavan} we show how the survival probability 
can be modified by this effect 
for the four cases of 
$(\alpha_{33},\alpha_{31}) = (\pm 5.7 \times 10 ^{-4}$ eV, 0), 
$(0, 2.5 \times 10 ^{-4}$ eV) and 
$(0, 2.5i \times 10 ^{-4}$ eV). 
As expected from the effective mass squared matrix in 
eq.~(\ref{eq:mass-matrix}), 
we can see in figure~\ref{fig-probab-mavan}, 
the effect of a nonzero $\alpha_{33}$ is to shift 
the position of the oscillation minimum, 
whereas that of $\alpha_{31}$ is to modify 
the oscillation amplitude (or effective 
mixing).

\section{Constraining New Physics Models}
\label{results-1}
In this section we present the sensitivity of a M\"ossbauer experiment
to constrain new physics based on the results of our $\chi^2$
analysis, described in the Appendix.

\subsection{Light Sterile Neutrino}
\label{steril}

\begin{figure}[h!]
\begin{center}
\hglue -0.1cm
\includegraphics[width=0.7\textwidth]{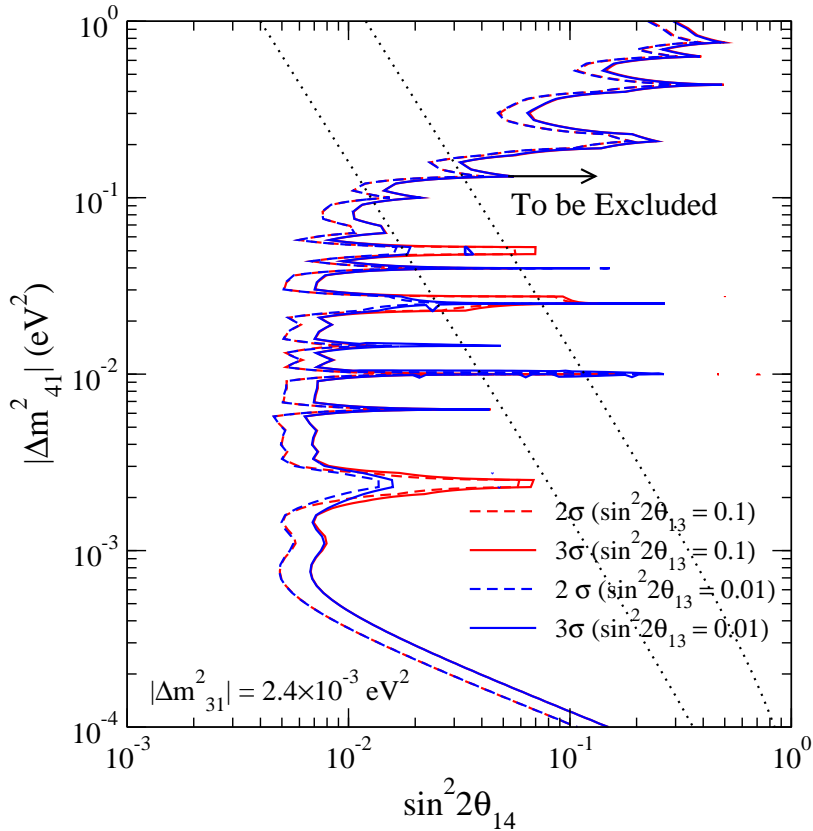}
\end{center}
\vglue -0.6cm
\caption{
Regions of the parameters $\Delta m^2_{41}$ 
and $\sin^2 2\theta_{14}$ which can be 
excluded if the data are consistent with 
the standard oscillation with 
$\sin^2 2\theta_{13} = 0.1$ by the red curves (see upper two lines of legend) 
or 0.01 by the blue curves (lower two lines of legend). 
The two dotted lines correspond to benchmarks of ref.~\cite{asrz}, 
see text.
}
\label{fig-steril-sens}
\end{figure}

In figure~\ref{fig-steril-sens} we show the region of the sterile neutrino 
mixing parameters that can be excluded if the data
are consistent with the standard three flavor active neutrino 
framework. 
%

We observe that the exclusion curves show somewhat 
complicated and strange oscilatory behaviours, 
leading to significant reductions of the sensitivity 
for particular values of $|\Delta m^2_{41}|$. 
We note that most of these behaviours are not physical 
and is caused by the fact that we have 
a finite number of detector positions. 
Below let us try to explain qualitatively the cause of such 
behaviours.

First of all, being a disappearance oscillation experiment, 
it is clear that  the standard oscillation driven by 
($\Delta m^2_{31}$, $\sin^2 2\theta_{13}$) without 
sterile neutrino, can always be mimicked by 
the same values of ($\Delta m^2_{41}$, $\sin^2 2\theta_{14}$) with 
vanishing $\theta_{13}$. 
This explains the ``dip'' behavior of the exclusion contours 
around $\Delta m^2_{41} = 2.4 \times 10^{-3}$ eV$^2$ 
in figure~\ref{fig-steril-sens}.
This loss of sensitivity can not be avoided even if 
we consider larger number of detector positions
(unless we use independent information from some other experiment). 

Second, for larger values of $|\Delta m^2_{41}|$ when
$\bar{\nu}_e$ survival probabilities exhibit many 
rapid oscillations, due to the finite number detector 
positions, there exist some special values of 
$|\Delta m^2_{41}|$ which reproduce quite well 
the original probabilities at all of the 
detector positions we considered. 
This is the cause of the loss of sensitivity 
at several particular values of 
$|\Delta m^2_{41}|$ larger than $\sim 5 \times 10^{-3}$ eV$^2$
see in figure~\ref{fig-steril-sens}.
We note, however, that in principle, by increasing the number 
of detector positions, such a loss of sensitivity can be avoided.

We also show in figure~\ref{fig-steril-sens} two dotted lines that 
correspond to benchmarks for the sterile neutrino contributions to the 
active neutrinos mass matrix. According to ref.~\cite{asrz}, if the 
oscillation parameters lie above the lower (upper) dotted lines, the sterile 
neutrino can influence sub-leading structures in the degenerate (normal 
hierarchy) neutrino mass spectrum.
We observe that the M\"ossbauer experiment can exclude a large region 
in the plane $\sin^2 2\theta_{14}$ versus $\vert \Delta m^2_{41}\vert$, in 
particular, it closes the lower mass window for any significant induced 
effect of the sterile neutrino in the active neutrinos mass matrix for 
the $\nu_e$-$\nu_s$ channel~\cite{asrz}.

\subsection{Large Extra Dimensions}
\label{LED}

\begin{figure}[h!]
\begin{center}
\includegraphics[width=0.7\textwidth]{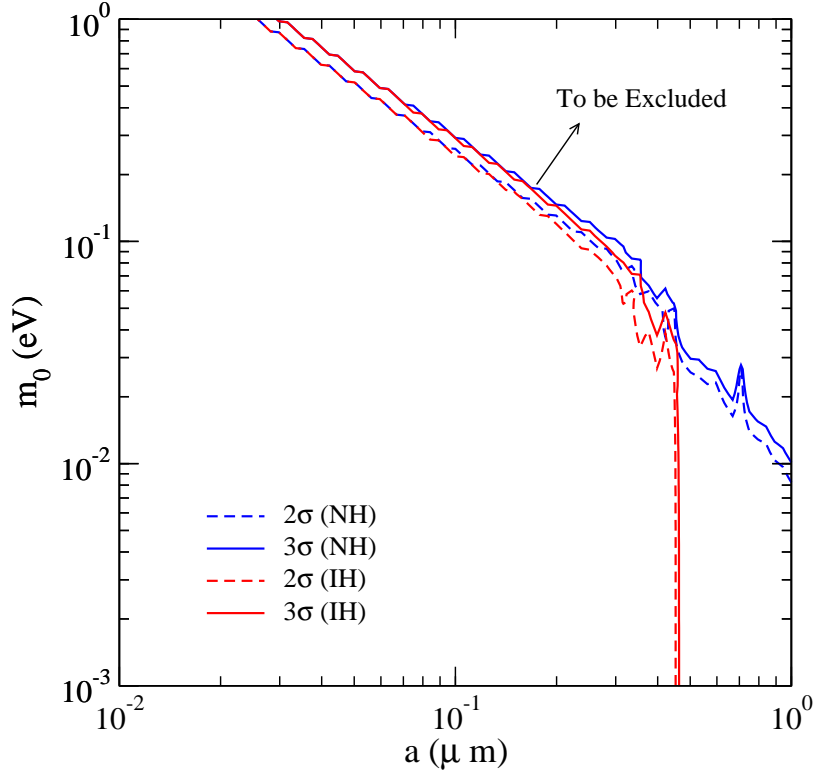}
\end{center}
\vglue -0.55cm
\caption{Region of parameters $m_0$ (lightest neutrino mass) and size
  of the largest extra dimension $a$ which can be excluded if the data
  are consistent with the standard oscillation (including the case for
  vanishing $\theta_{13}$).  The data was simulated with $\sin^2
  2\theta_{13}=0.1$ (as input) but the results do not essentially
  depend on the exact value of $\theta_{13}$.  }
\label{fig-sens-led}
\end{figure}

In figure~\ref{fig-sens-led} we show the sensitivity region in the LED
parameter plane $a$-$m_0$. Here  $a$ is the size of the largest 
extra dimension and $m_0$ is the lightest neutrino mass, 
$m_0$ being $m_1 (m_3)$ for the normal (inverted) mass hierarchy. 
The parameter region on the top-right side of the curves 
can be excluded by M\"ossbauer neutrinos if the data are 
consistent with the standard oscillation (including the case 
where $\theta_{13} = 0$). 

In obtaining these regions, we have also varied freely 
the LED parameters $a$ and $m_0$  and 
as described in the Appendix we have taken into account 
the finite size of the source and detector by using 
the Gaussian smearing function given in eq.~(\ref{eq:gaussian})
with $\sigma_L$ = 10 cm.
As expected from figure~\ref{fig-probab-led},
we obtained better sensitivity for the case of 
the inverted mass hierarchy. 
The value $\sin^22\theta_{13} = 0.1$ was used 
as input but we verified that in practice the results
do not depend on the true value of $\theta_{13}$. 

Compared to the current bound coming from CHOOZ, KamLAND and MINOS
obtained in \cite{Machado:2011jt}, the sensitivity we 
obtained here is somewhat better but very similar 
to the one which is expected 
from the Double CHOOZ  experiment \cite{Machado:2011jt}. 

%
We observe that there are  mainly two factors that reduce 
significantly the sensitivity of the M\"ossbauer experiment to LED.
 First, as mentioned in the previous section, despite the 
ultra monochromatic beam energy, due to the finite size of the 
source and detector, the LED effect which exhibit large
oscillatory behavior, is averaged out and the net effect
is significantly reduced. Second, the rather large correlated systematic 
uncertainty of 10\% which we assumed
following~\cite{Minakata:2006ne}
also reduces the sensitivity significantly for this model.

\subsection{Nonstandard quantum decoherence}
\label{decoherence}

\begin{figure}[h!]
\begin{center}
\hglue -0.1cm
\includegraphics[width=0.7\textwidth]{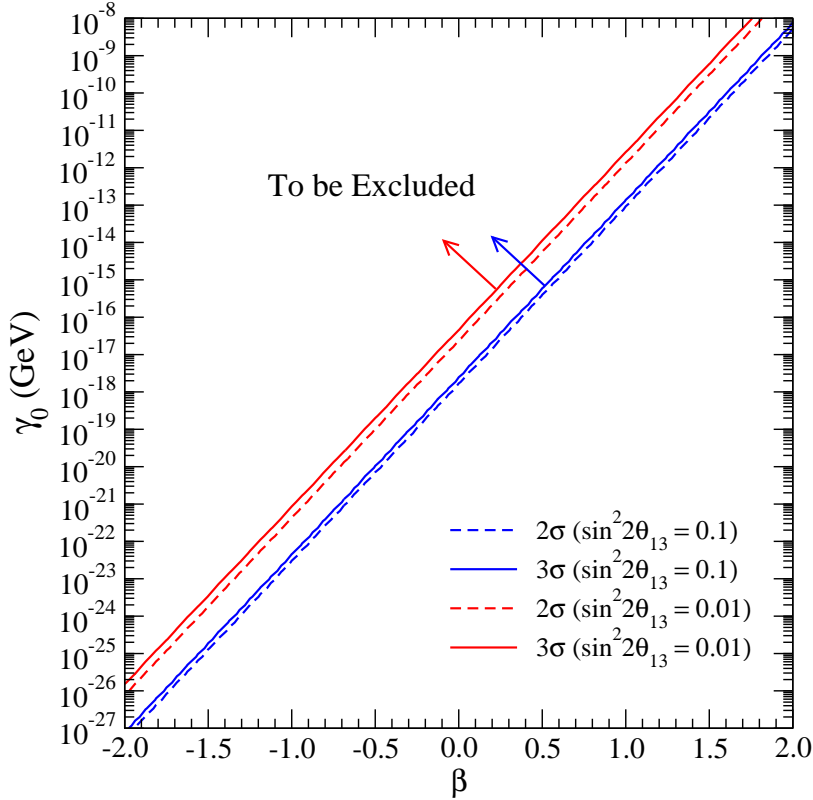}
\end{center}
\vglue -0.6cm
\caption{
Regions of the nonstandard decoherence 
parameters $\gamma_0$ and $\beta$ which can be 
excluded if the data are consistent with 
the standard oscillation and  
$\sin^2 2\theta_{13} = 0.1$ (upper two lines) 
or 0.01 (lower two lines). 
The region above the diagonal lines can be excluded 
(or probed) by the M\"ossbauer neutrino experiment.}
\label{fig-decoherence-sens}
\end{figure}

We show in figure~\ref{fig-decoherence-sens} 
the sensitivity regions in the plane of 
the NQD parameters $\beta-\gamma_0$
for the cases where the true value of the standard
parameter $\sin^2 2\theta_{13} = 0.1$ 
(lower two lines) or 0.01 (upper two lines).
The NQD parameters $\beta$ and $\gamma_0$, as well as 
the standard $\theta_{13}$ and $|\Delta m^2_{31}|$
were varied freely in fitting the data according to 
what is described in the Appendix. 
We find that the values of $\beta$ and $\gamma_0$ that lie above the
diagonal lines of figure~\ref{fig-decoherence-sens} are not compatible
with the simulated data and can be excluded by the M\"ossbauer
neutrino experiment, if the data are consistent with the standard
oscillation. On the other hand, the M\"ossbauer experiment has the
potential of observing NQD effects in this region.
Unlike the case of LED, the sensitivity to NQD essentially does not
depend on the mass hierarchy but strongly depends on the true value of
$\theta_{13}$. This can be easily understood from the expression of
the probability shown in eq.~(\ref{eq:prob-dc}).

We note that these results are worse,  
by $\sim$ 2-3 orders of magnitudes, than the 
current bounds on NQD obtained in ref.~\cite{Fogli:2007tx} 
which used solar and KamLAND neutrino data 
(see figure 1 of this reference). 
However, the bounds obtained in \cite{Fogli:2007tx} 
can not be compared directly to the results we 
obtained in this work because what was constrained by solar and 
KamLAND data was the decoherence effect relevant 
for oscillation between the first and second generation 
whereas we consider here the one between the first and the third generation. 
This has not yet been constrained for small $\theta_{13}$
allowed by current data, see \cite{Gago:2000qc}.

\subsection{Mass Varying Neutrinos}
\label{MaVaN}

\begin{figure}[h!]
\begin{center}
\hglue -0.2cm
\includegraphics[width=0.7\textwidth]{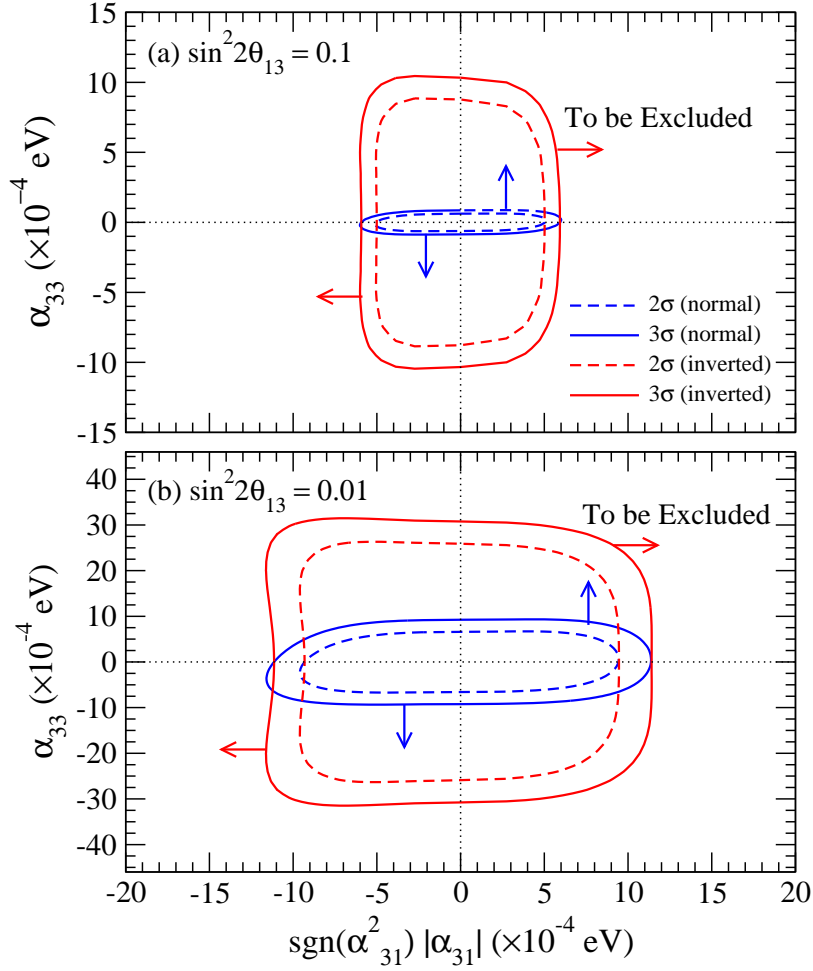}
\end{center}
\vglue -0.55cm
\caption{
Regions of MaVaN parameters $\alpha_{13}$ and $\alpha_{33}$ 
(outside the closed curves)
which can be excluded if the data are consistent 
with the standard oscillation 
with $\sin^2 2\theta_{13}$ = 0.1 (upper panel) or 
$\sin^2 2\theta_{13}$ = 0.01 (lower panel).}
\label{fig-sens-regions-MaVaN}
\end{figure}

In figure~\ref{fig-sens-regions-MaVaN}
we show the sensitivity regions of the MaVaN 
parameters of $\alpha_{13}$ and $\alpha_{33}$. 
Again we consider two possible input values for $\theta_{13}$, 
$\sin^2 2\theta_{13}$ = 0.1 (upper panel) or
$\sin^2 2\theta_{13}$ = 0.01 (lower panel). 
For simplicity, as in ref.~\cite{GonzalezGarcia:2005xu},
we have considered the case where CP violation 
is absent, so $\sin(\text{arg}[\alpha_{13}]) = 0$ ($\alpha_{13}$
is real or pure imaginary).
In order to determine these exclusion (sensitivity) regions, 
we have combined the results from 
two cases where the data are taken by inserting the matter 
between the source and the detectors (we assume the iron with $\rho = 7.9$ g/cm$^3$) 
and without matter which is considered practically as vacuum, 
which is crucial to constrain any MaVaN induced effect 
(comparison of these two cases is very important).  

We can exclude the parameter region outside the closed 
contours of figure~\ref{fig-sens-regions-MaVaN} for normal (blue lines) and 
inverted (red lines) mass hierarchy, if the data are consistent with standard 
oscillation. 
However if this is not the case, the M\"ossbauer experiment has the
potential to discover MaVaN effects in this region.

For the case where the mass hierarchy is normal, 
if $\sin^2 2\theta_{13} = 0.1$ (0.01), a 
M\"ossbauer neutrino experiment can exclude 
$|\alpha_{31}|$
$\gsim 6 \times 10^{-4}$ $(12\times 10^{-4}$) eV 
and 
$|\alpha_{33}|$ 
$\gsim 10^{-4}$ (10 $\times 10^{-4}$) eV
at 3 $\sigma$.
For the case where the mass hierarchy is inverted,
if $\sin^2 2\theta_{13} = 0.1$ (0.01), we can exclude 
the same range of $|\alpha_{31}|$ as in the case of normal hierarchy 
and 
$|\alpha_{33}|$  $\gsim 10^{-3}$ $(3\times 10^{-3}$) eV
at 3 $\sigma$.

Following \cite{GonzalezGarcia:2005xu}, one can try to 
describe the bounds we obtained in terms 
of more fundamental MaVaN parameters 
using eq.~(\ref{eq:scalar-coupling}). 
We conclude that, roughly speaking, 
for the normal mass hierarchy, 
if $\sin^2 2 \theta_{13} = 0.1$ (0.01) 
the ranges of 
\begin{equation}
| \lambda^\nu \lambda^f | \left( \frac{10^{-7} \text{eV}}{m_S} \right)^2
\gsim \times 10^{-27}\  (10^{-26}), 
\end{equation}
can be excluded by the M\"ossbauer experiment. 
For the inverted mass hierarchy, the bounds would 
be about one order of magnitude weaker. 

Comparing our results with the ones obtained in 
\cite{GonzalezGarcia:2005xu} which used solar neutrino data, 
the sensitivity we obtained is worse by 
a factor of $\sim 10$ or more. 
However, we should note that we are probing a 
different set of MaVaN parameters, relevant 
for the 1-3 sector which are not yet constrained by data. 

\section{Impact of New Physics on 
the determination of the standard parameters}
\label{results-2}

When one investigates the presence of any nonstandard property of
neutrinos, one also should worry about the impact these
new effects may have on the determination of the less known standard
mixing parameters.
In this section we discuss how the 
new physics studied in this paper 
can aggravate the determination of  
$\theta_{13}$ and $\vert\Delta m^2_{31}\vert$
in a M\"ossbauer neutrino oscillation experiment
(here we are mainly interested in the impact for 
$\theta_{13}$ since $\vert\Delta m^2_{31}\vert$ is 
already rather well determined.)

This can be easily achieved by projecting the four dimensional 
allowed parameter regions determined by our $\chi^2$ 
analysis described 
in the Appendix, into the plane of the standard mixing parameters
$\sin^2 2\theta_{13}$ and $|\Delta m^2_{31}|$. 

\begin{figure}[h!]
\begin{center}
\hglue -0.2cm
\includegraphics[width=0.74\textwidth]{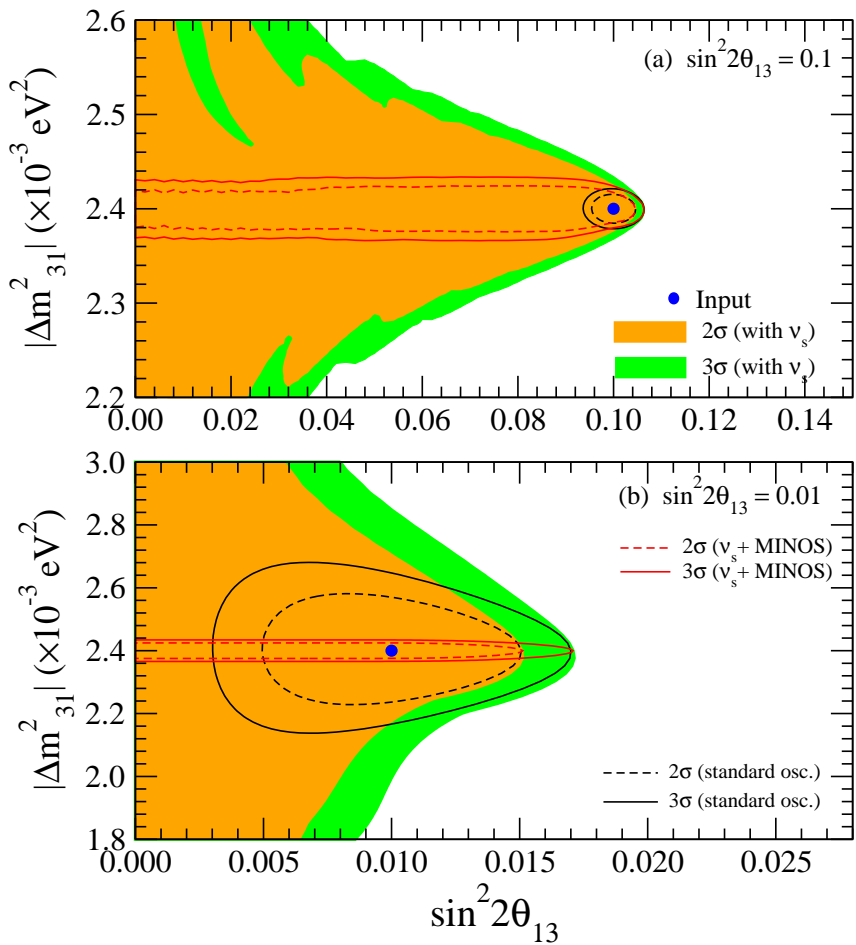}
\end{center}
\vglue -0.6cm
\caption{ Impact of the presence of a light sterile neutrino on the
  determination of $\theta_{13}$ and $|\Delta m^2_{31}|$.  We show the
  2 and 3 $\sigma$ CL regions allowed with (color shaded areas) and
  without (black solid and dashed curves) sterile neutrino parameters
  included in the fit.  Here the input values are $\sin^2 2\theta_{13}
  = 0.1$ (0.01) and $|\Delta m^2_{31}| = 2.4\times 10^{-3}$ eV$^2$.
We also show, by the red solid and dashed curves, 
the case where the information on the determination of 
$|\Delta m^2_{31}|$ from MINOS is combined. 
}
\label{fig-sin22q-sterile}
\end{figure}

When the presence of a light sterile neutrino is allowed in the fit 
we observe a large impact on the determination of the 
$\vert\Delta m^2_{31}\vert$ and $\theta_{13}$ oscillation parameters 
as can be seen in figure~\ref{fig-sin22q-sterile}
for the case where the true value of $\sin^2 2\theta_{13} = 0.1$
(upper panel) and 0.01 (lower panel).

In the case where $\vert\Delta m^2_{41}\vert$ and $\theta_{14}$ are varied
freely, the precise determination of $\vert\Delta m^2_{31}\vert$ and
$\theta_{13}$ by the M\"ossbauer experiment alone is severely limited.
Similar argument applies to the case of $\theta_{13}$ measurement by
reactor $\bar{\nu}_e$ alone.
This is because one can not identify if the reduction of the
$\bar{\nu}_e$ survival probability is due to nonzero $\theta_{13}$ or
nonzero $\theta_{14}$, with $|\Delta m^2_{41}|$ similar to 
$|\Delta m^2_{31}|$.  
In particular, for vanishing $\theta_{13}$, arbitrary value of
$|\Delta m^2_{31}|$ is allowed (see figure~\ref{fig-sin22q-sterile})
as the input can be easily 
mimicked by $\sin^2 2\theta_{14} \sim 0.1$ (0.01)
and $|\Delta m^2_{41}| \sim 2.4\times 10^{-3}$ eV$^2$.
Even if we add information on the allowed values of
$|\Delta m^2_{31}|$ from another experiment, like MINOS, 
one can not anymore determine the mixing angle, even if it 
is rather large, by the
M\"ossbauer experiment.

\begin{figure}[h!]
\begin{center}
\hglue -0.2cm
\includegraphics[width=0.74\textwidth]{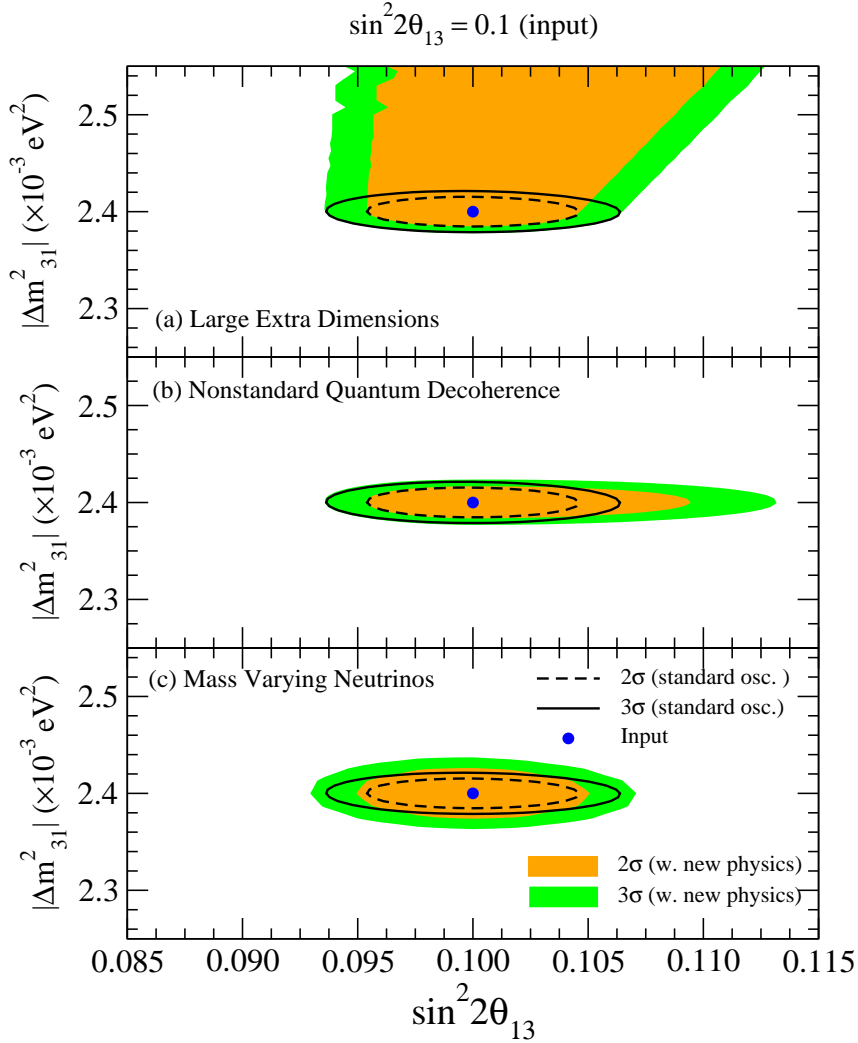}
\end{center}
\vglue -0.6cm
\caption{Impact of LED (upper panel), NQD (middle panel) and MaVaN
  (bottom panel) on the determination of the $\sin^2 2\theta_{13}$ and
  $|\Delta m^2_{31}|$.  We show the 2 and 3 $\sigma$ CL allowed
  parameter regions determined when the true (input) value is $\sin^2
  2\theta_{13} = 0.1$ (color shaded areas). For reference we also show
  the 2 and 3 $\sigma$ CL allowed regions without considering the
  possibility of new physics in the fit (solid and dashed lines). }
\label{fig-sin22q=0.1}
\end{figure}

\begin{figure}[h!]
\begin{center}
\hglue -0.2cm
\includegraphics[width=0.74\textwidth]{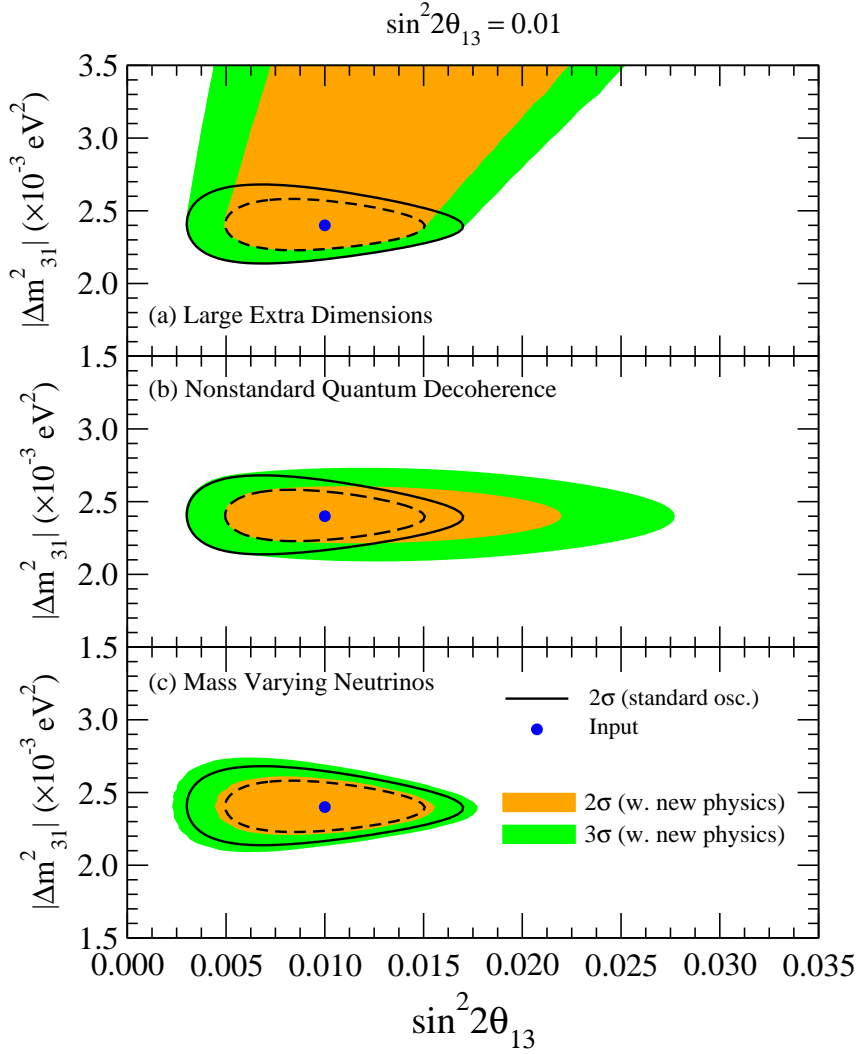}
\end{center}
\vglue -0.6cm
\caption{
Same as in figure~\ref{fig-sin22q=0.1}
but when the true value is $\sin^2 2\theta_{13} = 0.01$.}
\label{fig-sin22q=0.01}
\end{figure}

In Figs. \ref{fig-sin22q=0.1} and \ref{fig-sin22q=0.01}, for the case
of $\sin^22\theta_{13} = 0.1$ and 0.01, respectively, we show how the
presence of LED (upper panel), NQD (middle panel) and MaVaN (lower
panel) can influence the determination of the standard mixing
parameters.

For LED one loses sensitivity in the determination of both 
$\vert \Delta m^2_{31}\vert$ and $\sin^2 2\theta_{13}$
but only towards the larger values of these parameters. 
This is because in the limit of small LED effects, 
consistent with current experimental bounds, 
one can write~\cite{Davoudiasl:2002fq,Machado:2011kt} the effect 
of LED in terms of the effective mass squared difference 
$$ \Delta {m^2_{31}}^{\text{(eff)}} \approx \Delta m^2_{31} - \frac{\pi^2}{3}\, a^2 \; \Delta m^4_{31} \, ,$$
and the mixing angle 
$$ \sin^2 \theta_{13}^{\text{(eff)}} \approx \sin^2 \theta_{13} \, (1-
\frac{\pi^2}{3} \, m_3^2 a^2)\, ,$$ so that bigger values of $\vert
\Delta m^2_{31} \vert$ can fit the data as long as they can be
compensated by a corresponding increase of size of the largest extra
dimension $a$.  At the same time when one increases $a$, one decreases
$\sin^2 \theta_{13}^{\text{(eff)}}$ so one needs to increase
$\sin^2\theta_{13}$ in order to fit the data.  Also because of the
above behavior the minimum values of $\vert \Delta m^2_{31}\vert$ and
$\sin^2 \theta_{13}$ allowed by LED coincide with the ones allowed by
the standard analysis.

For the case of NQD, the allowed parameter regions distorted only
towards larger values of the mixing angle. This is because the net
effect of NQD is to reduce the amplitude and therefore, NQD can be
compensated (canceled) to some extent by a larger value of the mixing
angle.  We conclude that, NQD, if present, could induce a significant
overestimation of the mixing angle $\theta_{13}$.

For the MaVaN model we considered in this work, 
in principle, we do not have any problem in determining 
the standard mixing parameters because one can remove the 
matter inserted between source and detector for this determination.
The case where only the atmosphere is 
present can be regarded as vacuum as the density 
of the atmosphere is too small to induce any MaVaN effect. 
However, for the sake of discussion we present the results for the
case where the experiment is performed with and without matter (iron)
and combined.

From the bottom panels of 
figure~\ref{fig-sin22q=0.1} and \ref{fig-sin22q=0.01}, 
we can see that the allowed parameter region for  
$|\Delta m^2_{31}|$ and $\theta_{13}$ are slightly 
increased, which was expected, 
since the parameters $\alpha_{33}$ and $\alpha_{31}$
can mimic the mass squared difference and the mixing
angle, respectively, as we can see from eq.~(\ref{eq:mass-matrix}).  
We conclude that even in this case the impact of MaVaN in the determination 
of $|\Delta m^2_{31}|$ and $\theta_{13}$ is small.

\section{Discussion and Conclusions}
\label{sec:conclusions}

In this work, we discussed the potential of a 
short baseline ($\sim O(10\ \text{m})$) M\"ossbauer neutrino 
oscillation experiment based on the $^3$H-$^3$He system
to probe new physics beyond the standard model.
We investigate four different scenarios: the presence of 
a light sterile neutrino that can mix with $\bar \nu_e$, 
a model where a tower of sterile neutrinos can change the $\bar \nu_e$
oscillation pattern due to large flat extra dimensions, 
neutrino oscillations with nonstandard quantum decoherence 
and mass varying neutrinos.

When a single light sterile neutrino is added to the standard three
active flavor neutrinos, we conclude that a M\"ossbauer oscillation
experiments can probe (exclude) parameter regions still not yet
excluded by other experiments. In particular, it can close the lower
mass window for any significant induced effect of the sterile in the
active neutrino mass matrix for the $\nu_e-\nu_s$ channel~\cite{asrz}.

For the LED model, M\"ossbauer neutrinos can exclude the size of the
largest extra dimension $a \gsim 1\ (0.45)$ $\mu$m at 3 $\sigma$ CL
for the normal (inverted) mass hierarchy for a vanishing lightest
neutrino mass ($m_0=0$).  If $m_0$ is larger, say 0.2 eV, for example,
this experiment can exclude $a \gsim 0.15$ $\mu$m at 3 $\sigma$ CL.
The sensitivity we obtained is somewhat better than the current bounds
but not better than what can be achieved in the near future by reactor
experiments such as Double CHOOZ.  We note, however, the sensitivity
of the M\"ossbauer experiment can be improved if one can reduce the
uncertainties on the neutrino production and detection positions as
well as the correlated systematic uncertainty on the initial neutrino
flux and/or capture cross section.

For NQD, due to the low energy, M\"ossbauer neutrinos are most
sensitive to the case of $\beta = -2$, where $\gamma_0 \gsim 10^{-27}$
($ 10^{-26}$) GeV can be excluded if $\sin^22 \theta_{13} = 0.1$
(0.01).  If $\beta$ is increased by a unity, the sensitivity would be
reduced by about a factor of five orders of magnitude.  The
sensitivity we obtained for the NQD
parameter are worse than the existing ones derived from  solar and
KamLAND neutrino data~\cite{Fogli:2007tx}.  
However, the existing bounds are for the mixing 
between first and second generation whereas M\"ossbauer neutrinos can put
limit on the decoherence parameter for first and third generation, 
where no bounds currently exist (for small $\theta_{13}$.)

Regarding the MaVaN model, for $\sin^2 2\theta_{13} = 0.01-0.1$,
M\"ossbauer neutrinos can exclude the range of $|\alpha_{31}|$ and
$|\alpha_{33}|$ parameters larger than $\sim 10^{-4}-10^{-3}$ eV,
depending on the precise value of $\theta_{13}$ and of the mass
hierarchy.  This is worse than the bounds obtained in
\cite{GonzalezGarcia:2005xu} which used solar neutrino and KamLAND
data but as in the case of the decoherence effect, the existing bounds
apply only to the 1-2 sector whereas M\"ossbauer neutrinos can probe
the 1-3 sector which is not bounded yet.

We still do not know if a M\"ossbauer neutrino oscillation
experiment can be really feasible due to several technical
(experimental) difficulties. Nevertheless, we hope that new
technologies will come about permitting it to happen in the future.
It would be fantastic to be able to build such a compact experiment
capable to explore not only standard but also nonstandard neutrino
oscillation physics.

\section*{Acknowledgments}
This work is supported by Funda\c{c}\~ao de Amparo \`a Pesquisa do
Estado de S\~ao Paulo (FAPESP), Funda\c{c}\~ao de Amparo \`a Pesquisa
do Estado do Rio de Janeiro (FAPERJ) and by Conselho Nacional de
Ci\^encia e Tecnologia (CNPq). The work of PANM has also been
supported by an European Commission ESR Fellowship under the contract
PITN-GA-2009-237920.  Three of us (PANM, HN, RZF) would like to
acknowledge the Fermilab Theory Group for its hospitality during the
last stage of this work.

\appendix 
\section{Analysis Method}
\label{appendix-analysis-method}


Here we give a very brief description of our analysis method
which is basically the same as used in \cite{Minakata:2006ne,Minakata:2007tn}
apart from the fact that we have more free parameters due to new physics. 

We adopt the experimental setup referred to as Run IIB in 
\cite{Minakata:2006ne} where the detectors occupy 10 different 
positions corresponding to the following baselines:
$L_1 = L_{\text{OM}}/5$, $L_{i+1} = L_i + (2/5)L_{\text{OM}}$, $i = 1,..., 9$, 
where  $L_{\text{OM}} \equiv 4\pi E/|\Delta m^2_{31}| \simeq 9$ m,
is the distance which corresponds to the first oscillation minimum. 
We assume that each detector is exposed to 10$^6$ $\bar{\nu}_e$ events. 

We simulate the input data assuming only standard oscillation physics.
In doing this, throughout this work, we assume that the true values
(input values for the simulation) of the standard mixing parameters
are $\Delta m_{21}^2 = 7.6 \times 10^{-5}$ eV$^2$, $\sin^2
\theta_{12}$ = 0.31, $|\Delta m_{31}^2| = 2.4 \times 10^{-3}$ eV$^2$,
and $\sin^2 2\theta_{13}$ = 0.1 or 0.01.
In our $\chi^2$ analysis (described below), when fitting the simulated
data, we vary freely, in addition to the new physics parameters of
each model, the standard mixing parameters $\theta_{13}$ and $|\Delta
m_{31}^2|$.  We do not vary the other oscillation parameters since the
impact of their uncertainties is quite small.  In order to illustrate
the standalone potential of this experiment, we do not use any biases
information on the values of $|\Delta m^2_{31}|$ and $\theta_{13}$
from existing or future experiments, 
except for results shown in figure~\ref{fig-sin22q-sterile}
where the results from MINOS experiment is combined 
only for the purpose of illustration of loss of 
the sensitivity. 

When we calculate the survival probability for the LED model
we observed many rapid oscillations due to the conversion
of $\bar{\nu}_e$ into KK modes, as can be seen in 
figure~\ref{fig-probab-led} in section~\ref{LED-framework}.
Due to the finite size of the source and detector
(uncertainties on the exact production and detection 
positions), such rapid oscillations must be 
averaged out over the size of the source and detector, 
which tends to ``wash out'' the LED effect. 
In order to take this finite size effect
into account,  we average the probability over 
the baseline, allowing for the uncertainty on the production/detection 
points through a Gaussian smearing function defined as 
\begin{equation}
f(L,L') = \frac{1}{\sqrt{2\pi} \sigma_L} 
\exp\left[-\frac{(L-L')^2}{2\sigma_L^2} \right],
\label{eq:gaussian}
\end{equation}
where we set $\sigma_L$ = 10 cm.

To evaluate the sensitivity to constrain new physics
described by the parameters, say, $\alpha$ e $\beta$, 
we compute 
\begin{equation}
\Delta \chi^2_{\text{min}}(\alpha, \beta) 
= \chi^2_{\text{min}}(\alpha, \beta)
-\chi^2_{\text min}(\alpha= \beta = 0)\, ,
\end{equation}
where 
$\chi^2_{\text {min}}(\alpha,\beta)$ is the minimum 
of 
$\chi^2(\alpha,\beta) \equiv 
\chi^2(\alpha,\beta,\Delta m^2_{31}, \sin^2 2\theta_{13})$ given by 
\begin{eqnarray}
& & \chi^2(\alpha, \beta, 
\Delta m^2_{31},\, \sin^2 2\theta_{13}) 
= 
 \sum_{i,j=1}^{10} 
\left[
\frac{ N_i^{\text{obs}}-N_{i}^{\text{theo}}}{N_i^{\text{theo}}}\right]\, 
(V^{-1})_{ij}\, 
\left[
\frac{N_j^{\text{obs}}-N_{j}^{\text{theo}}}{N_j^{\text{theo}}}\right]
\, ,
\label{eq:chi2}
\end{eqnarray}
%
where $N_i^{\text{obs}}$ is the number of observed (simulated) events at
baseline $L_i$ for a given fixed values of the standard
oscillation parameters (no new physics), while $N_i^{\text{theo}} =
N_i^{\text{theo}}(\alpha, \beta, \theta_{13}, \Delta m^2_{31})$ is the
theoretically expected number of events at baseline $L_i$ for a given
set of oscillation and new physics parameters.
We use, as in ref.~\cite{Minakata:2006ne}, 
the correlation matrix defined by the elements
\begin{equation}
(V^{-1})_{ij}= \frac{\delta_{ij}}{\sigma^2_{ui}} - 
\frac{1}{\sigma^2_{ui}\sigma^2_{uj}}
\frac{\sigma^2_c}{[1+(\sum_{k} \frac{1}{\sigma^2_{uk}})\sigma^2_c]},
\end{equation}
where 
$\sigma^2_{ui} = \sigma^2_{\text{usys}} + 1 / N_{i}^{\text{theo}}$. 
For the uncorrelated systematic uncertainty, 
we take the optimistic choice considered in\cite{Minakata:2006ne} 
$\sigma_{\text{usys}}=0.2$ \%.
For the correlated systematic uncertainty we set, 
as in\cite{Minakata:2006ne} , $\sigma_c=10$ \%.

We determine the new physics parameter regions that  
can be excluded by the M\"ossbauer experiment 
by imposing the condition $\Delta \chi^2_{\text{min}} > 6.18 $ 
and 11.83, respectively, for 2 and 3 $\sigma$ significance level
and 2 degrees of freedom.

\end{document}